\documentclass[preprint,showpacs]{revtex4-2}
\usepackage{graphicx}
\usepackage{amsmath}
\usepackage{amssymb}

\begin{document}

\title{Unification based on the mysterious cubic-structure grouping of quarks and leptons}

\author{Mohammad Mehrafarin}
\email{mehrafar@aut.ac.ir}
\affiliation{Physics Department, Amirkabir University of Technology, Tehran 15916, Iran}

\begin{abstract}
We present a unification model based on the well-known but mysterious cubic-structure grouping of quarks and leptons that suggests an underlying symmetry connection deemed explainable by a unified theory. It results in an extension of the Pati-Salam model that consolidates the fermions into two sixteen-dimensional chiral representations of the gauge group. Moreover, the discrete cubic flavor symmetry arbitrarily conjectured in the literature is found here to be implied by the gauge symmetry. Furthermore, the gauge algebra also describes an eleven-dimensional spacetime decomposed into the usual spacetime and the seven-sphere, which is the manifold of unit octonions. This suggests such a spacetime is apt for embracing all elementary particles and their interactions.
\end{abstract}
 

\maketitle

\section{Introduction}
At sufficiently small length scales, nature is described by indivisible elementary particles, which make up the matter and carry their interactions. Of the four fundamental interactions, the weak, the strong, and the electromagnetic forces are described by gauge theories of a similar kind, hinting that their differences may not be fundamental.  This pointed to the discovery of the electroweak (EW) unification of electromagnetism and the weak force \cite{Glashow1961,Salam1964,Weinberg1967}, and further alludes to a grander unification. A grand unified theory (GUT) is expected to provide a rationale for some established peculiarities of elementary particles and predict new phenomena that cannot be deduced from the Standard Model (SM) of particle physics. However, previous attempts at formulating a GUT (e.g., \cite{PS,SU5,SO10,E6,SU2n,SU15}) have not resulted in a generally accepted physical model.

In this article, we present a unification model based on the well-known but overlooked cubic structure grouping of leptons and quarks that is considered to be explainable by a unified theory \cite{Georgi}. This mysterious cubic structure is explicitly demonstrated in the next section.  Its two-dimensional projection on the leptonic diagonal forms the Star of David figure that is explained by the color rep ($\mathbf{1}\oplus\mathbf{1}\oplus\mathbf{3}\oplus{3}^*$) of the rank-two subalgebra $su(3)_C$ of SM (bold number denotes the dimension of a rep and star means dual). The cube takes shape by adding the third dimension to this figure via displacing each particle according to its electric or weak hypercharge. The fact that specific properties of leptons and quarks can be graphed in three dimensions by this simple structure hints at some symmetry connection, which this article explores.

The cubic grouping is implicit in, and partially explained by, the  Pati-Salam (PS) model \cite{PS} due to its rank-three symmetry subalgebra $su(4)$.  Regarding lepton number as the fourth color, the representation (rep) $\mathbf{4}\oplus \mathbf{4}^*$ of the Lie algebra  $su(4)$ can be visualized by the four color and the four anticolor states on the vertices of a cube (when restricted to its subalgebra $su(3)_C$, the rep decomposes as $\mathbf{1}\oplus\mathbf{1}\oplus\mathbf{3}\oplus{3}^*$, mentioned above). However, this rep is reducible and does not bring about the full connection/mixing between the states implicitly suggested by the cubic grouping. To fully explain the structure, we need a  larger {\it self-dual} algebra that can consolidate the octet of colors and anticolor states (the ``color'' octet) into an irreducible rep (irrep). We find that the Lie algebra $so(7)$ is the minimal extension of the symmetry $su(4) \cong so(6)$ that naturally unifies the color octet in its real eight-dimensional spinor rep $\mathbf{8}$. Thus, the gauge symmetry group of our unification model is $SU(2)_L\times SU(2)_R\times Spin(7)$, which extends the PS model and contains six additional (confined) gauge bosons, further mixing the fermions. (Other extensions of PS symmetry, in such a way as to incorporate new physics in the Tev region, have been considered before \cite {Foot}.) The group $Spin(7)$ is real, consistent with the fact that it unites colors and anticolors in the same irrep. It yields a grouping of the thirty two fermions of a given generation into two sixteen dimensional irreps $(\mathbf{2}\otimes\mathbf{1}\otimes\mathbf{8})$ and $ (\mathbf{1}\otimes\mathbf{2} \otimes\mathbf{8}) $ pertaining to the left chiral (LC) and right chiral (RC) fermions (by fermions we collectively refer to both fermions and anti-fermions, of course).

Moreover, we show that $so(7)$ entails the discrete cubic symmetry group $S_4$, which re-establishes the flavor grouping in the cubic structure. Therefore, the flavor symmetry $S_4$ {\it conjectured} in the literature independently of the gauge groups  (e.g., \cite{Ma2004,Hagedorn2006,SO10S4,Ishimori,MA2006,flavor}), here is implied by the gauge symmetry. They both come in one package. This discrete symmetry has led to the quark-lepton complementary relation as well as realistic mixing matrices and mass hierarchies \cite{MA2006,flavor,mixing,mixing2}.

Furthermore, we show that the algebra of the gauge group also describes the group  $SO^+ (3,1)\times Spin(7)$ representing the symmetry of the eleven-dimensional spacetime $M^4 \times S^7$, where $M^4$ is the usual four-dimensional spacetime and the $S^7$ is the seven-sphere. The latter is the manifold of unit octonions whose tangent space at each point, spanned by imaginary octonions,  has the rotation group $Spin (7)$. The isomorphism in algebra suggests that the manifold  $M^4 \times S^7$ is apt for embracing all elementary particles and their interactions. Thence, the gauge group $Spin(7)$ is also a spacetime symmetry group, and its spinor rep $\mathbf{8}$ which describes the color states, is a $S^7$ spinor. The size of the seven-sphere is determined by the energy scale where the $Spin(7)$ symmetry breaks.

More detailed investigations of the proposed model can include, for example, the study of the lepto-quark mixing with an eye on realistic mass hierarchies and new sources of charge-parity violation, as well as the physics connected with the seven-sphere manifold \cite{Duff,Ranjbar,Nilsson} and its implication for octonionic description of elementary particles \cite{Krasnov,Singh,Furey}. 

\section{The cubic grouping in the PS model}\label{PS}
In the PS model, with gauge group $SU(2)_L \times SU(2)_R \times SU(4)$, color and $B-L$ hypercharge are unified in $SU(4)$. The $B-L$ hypercharge is a part of the weak hypercharge, the other part being in $ SU(2)_R$. With the onset of parity symmetry breaking, the two groups break as
$ SU(2)_R\times SU(4) \rightarrow SU(3)_C \times U(1)_Y$
so that the two parts combine to form the SM weak hypercharge $U(1)_Y$. As for the electric charge, part of it descends from $SU(2)_R$ to  $U(1)
_Y$, while the other is in $SU(2)_L$. In the EW symmetry breaking, the latter two groups break as $SU(2)_L \times U(1)_Y \rightarrow U(1)_Q$, combining the two parts to give the electric charge.
\begin{figure}
\centering
{\includegraphics[width=3.3cm]{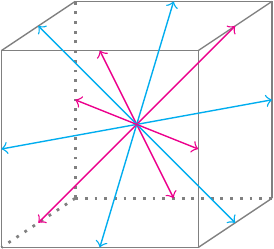}}\\
\begin{tabular}{ccc}
{\includegraphics[width=4 cm]{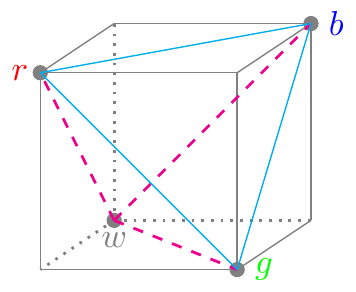}}&\quad
{\includegraphics[width=4 cm]{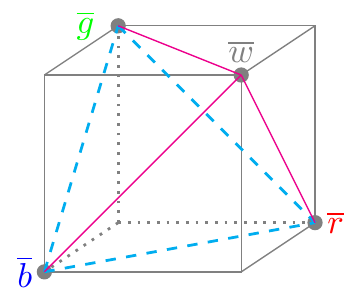}}&\quad
{\includegraphics[width=4 cm]{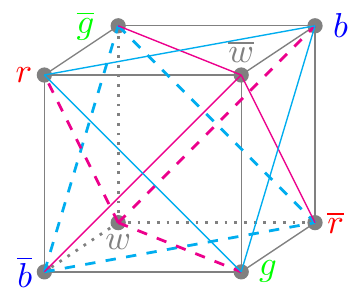}}
\end{tabular}
\caption{Representations of the Lie algebra $su(4)$. The linear dimensions of the top cube are twice the dimensions of the bottom cubes (drawn not to scale). The adjoint rep $\mathbf{15}$ (top), has twelve root vectors whose ends lie on the midpoints of the edges. The cyan vectors are the roots of the subalgebra $su(3)$. The standard rep $\mathbf{4}$ (bottom left) describing the four color states and the dual rep $\mathbf{4}^*$ (bottom middle)  describing the four anticolor states,   with root vectors connecting the weights, are shown within the reference cube. The self-dual rep  $\mathbf{4}\oplus\mathbf{4}^*$ (bottom right) describes the color octet. The cyan triangles in $\mathbf{4}$ and $\mathbf{4}^*$, respectively, correspond to irreps $\mathbf{3}$ and $\mathbf{3}^*$ of the subalgebra $su(3)$.}\label{fig1}
\end{figure}

The cubic-structure grouping of fermions is implicit in the PS model, partially being explained by the subalgebra $su(4)$. Let us make this point explicit,  which puts the model in a perspective that naturally leads to a grander unification. The four color states  $r$ (red), $g$ (green), $b$ (blue), $w$ (white), where $w$ pertains to leptons, form the standard rep $\mathbf{4}$ of the algebra $su(4)$. When restricted to $su(3)_{C}$, we have $\mathbf{4}=\mathbf{1} \oplus \mathbf{3}$. Moreover, the anticolor states are represented by the dual rep $\mathbf{4}^*$, and the color octet by the self-dual rep $\mathbf{4}\oplus \mathbf{4}^*$. Having rank three, these representations of $su(4)$ form a cube \cite{Fulton}, as shown in Fig.~\ref {fig1}. To elaborate on this figure, we informally state a few preliminary facts about the representation theory of semisimple algebras: (i) The maximal set of commuting generators of an algebra is the Cartan subalgebra, whose dimension is the rank of the algebra. (ii) In a given irrep, the common eigenvectors of the Cartan subalgebra, labeled by their eigenvalues, uniquely characterize the irrep. (iii) These labels define points (vectors), called weights (or weight vectors), in the space of dimension equal to the rank, called the weight space. Thus, the weights characterize the rep. (iv) The weights of the adjoint rep are specified as roots, and its space, the root space. This is because one can construct all other reps from the root system. (v) Root vectors always come in opposite pairs, and zero vectors are always equal in number to the rank. The latter vectors are trivial and not explicitly included in the root system.  (vi) When the root system is applied to the weights of a given irrep, opposite root vectors, shown by the same undirected line, connect weights in pairs. 

\begin{figure}
\centering
\begin{tabular}{cc}
{\includegraphics[width=4.5 cm]{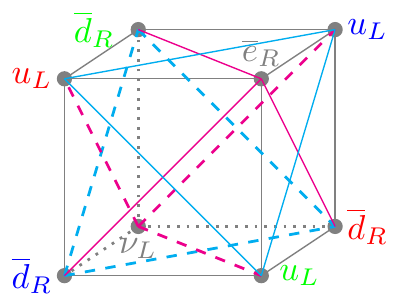}}&\qquad
{\includegraphics[width=4.5 cm]{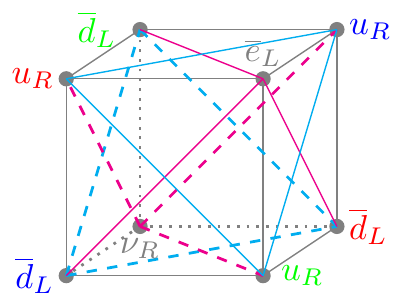}}\\
{\includegraphics[width=4.5 cm]{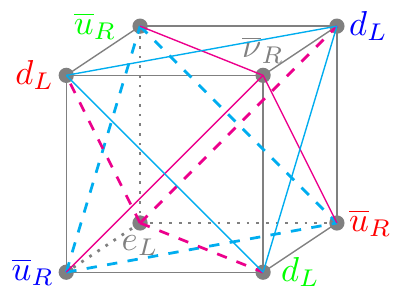}}&\qquad
{\includegraphics[width=4.5 cm]{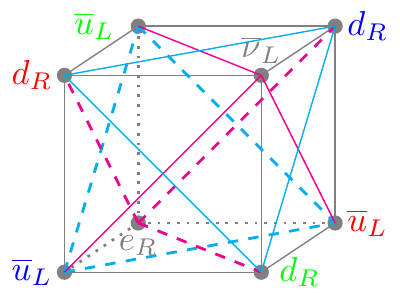}}
\end{tabular}
\caption{Fermion cubic grouping. Groupings of the thirty two fermions of a given generation into two LC (left) and two RC (right) cubic structures represented, respectively, by $\mathbf{2}\otimes \mathbf{1}\otimes(\mathbf{4}\oplus\mathbf{4}^*)$ and $\mathbf{1} \otimes \mathbf{2}\otimes(\mathbf{4}\oplus\mathbf{4}^*)$. Weights correspond to fermions, while the cyan and magenta root vectors are the intermediating $Y$ gauge bosons and gluons, respectively.  ($\nu, e, u, d$ are generic, belonging to any generation.)}\label{fig2}
\end{figure}

The space $\mathbf{2}\otimes\mathbf{1}$ (resp. $\mathbf{1}\otimes\mathbf{2}$) of the standard rep of $SU(2)_L$ (resp. $SU(2)_R$) is spanned by the left (resp. right) isospin up/down states. They have no color and only correspond to isospin; color comes from the space of the color octet, which is spanned by the four color and four anticolor states. Thus, for example, tensoring $w$ with the left isospin-up (resp. down) state gives $\nu_L$ (resp. $e_L$). Therefore, by tensoring the color octet visualized in Fig.~\ref{fig1} with  left and right isospin up/down states, we obtain a grouping of the thirty two fermions of a given generation into two LC and two RC cubes reducibly represented by  $\mathbf{2}\otimes \mathbf{1}\otimes(\mathbf{4}\oplus\mathbf{4}^*)$ and $\mathbf{1} \otimes \mathbf{2}\otimes(\mathbf{4}\oplus\mathbf{4}^*)$, as shown in Fig.~\ref{fig2}.

Physically, the weights of a rep correspond to fermions while the roots represent intermediary gauge bosons between them. An undirected line representing two opposite roots, thus, represents two oppositely charged gauge bosons mediating between a pair of fermions.  Gauge interactions that do not change the identity of a fermion correspond to zero vectors not shown in the root system. In the case of $su(4)$, they correspond to the $B-L$ hypercharge and two gluonic interactions. The cyan lines shown within each cube in Fig.~\ref{fig2} represent gluons ($su(3)_C$ root vectors) mediating the interconversion of corresponding quarks. The magenta lines mediate the interconversion of lepto-quark pairs of the same helicity by exchanging six additional gauge bosons predicted by the PS model. Collectively labeled $Y$, three of these bosons are seen to carry color charge $r, g$, or $b$, electric charge $\frac{2}{3}$, and weak hypercharge $\frac{4}{3}$, and the other three have opposite charges (all are thus confined like gluons). We denote the first three as $Y^+_{r}, Y^+_{g}, Y^+_{b}$ and the second three as  $Y^-_{\overline{r}}, Y^-_{\overline{g}}, Y^-_{\overline{b}}$. Therefore, the PS model contains fifteen gauge bosons in the $su(4)$ sector in accord with the dimension of the subalgebra. The $Y$ bosons yield charged current lepto-quark interactions. As a corollary of their electric charge and hypercharge values, these charges must be quantized for all particles as $\frac{1}{3}\mathbb{Z}$. 

The curious feature mentioned in the Introduction about the cubic grouping is explicitly apparent in Fig.~\ref{fig2}. By translating along the leptonic diagonal through the four flavor classes (which is the same as translation along the edges), the electric charge and the weak hypercharge shift respectively by $\frac{1}{3}$ and $\frac{2}{3}$. Considering the change in color charge, this hints at the interconversion of fermion pairs by the exchange of another six gauge bosons represented by the edges of the cube. These are not contained in the PS model, but we can incorporate them by enlarging the symmetry $su(4)$ of the color octet to $so(7)$, which at the same time unites the rep $\mathbf{4}\oplus\mathbf{4}^*$ into an irrep of the enlarged algebra. This would explain the full connection between the fermions implicitly implied by the cubic grouping.

\section{The gauge group of the unification model}
As the primary step in our unification scenario, we show that $su(4) \cong so(6)$ can be extended to the $so(7)$ symmetry algebra, which will consolidate the color octet into its real (self-dual) eight-dimensional spinor rep $\mathbf{8}$. It corresponds to the enlargement of the PS model gauge group to $SU(2)_L \times SU(2)_R \times Spin(7)$, introducing six additional intermediating gauge bosons to reconcile with the twenty-one dimensions of $so(7)$.

\begin{figure}
\centering
\begin{tabular}{ccc}
{\includegraphics[width=3.5cm]{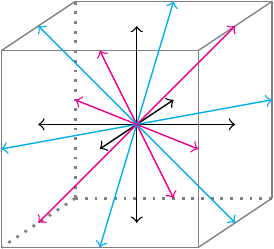}}&\qquad\quad
{\includegraphics[width=3.5cm]{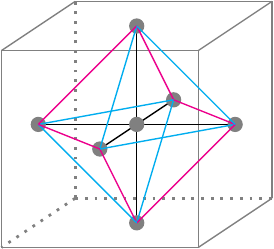}}&\qquad\quad
{\includegraphics[width=3.5cm]{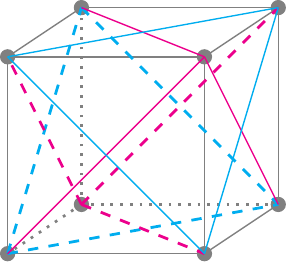}}
\end{tabular}
\caption{Fundamental representations of the Lie algebra $so(7)$. The linear dimensions of the first two cubes are twice the dimensions of the third (drawn not to scale). The adjoint rep (left) $\mathbf{21}$, with twelve long (cyan and magenta) and six short (black) root vectors of length ratio $\surd{2}$, are shown within the reference cube. The ends of the root vectors lie on the midpoints of the edges and faces of the cube. The magenta and cyan vectors are the roots of the subalgebra $su(4)$, already seen in Fig.~\ref{fig1}. The black vectors are the additional new roots. The standard rep $\mathbf{7}$ (middle) and the spinor rep $\mathbf{8}$ (right) are also shown with the roots connecting the weights.}\label{fig3}
\end{figure}

The Lie algebra $so(7)$ of $Spin(7)$ also has rank 3, compatible with the cubic structure. Its fundamental reps are depicted in Fig.~\ref{fig3} \cite{Fulton}.
\noindent When restricted to its subalgebra $so(6)$, the spinor rep $\mathbf{8}$ decomposes as $\mathbf{4}\oplus\mathbf{4}^*$  which is the color octet of Fig.~\ref{fig1}. Therefore,  the octet can be described by irrep $\mathbf{8}$ by extending its symmetry, thus yielding six additional roots, which are the black roots along its edges seen in Fig.~\ref{fig3}. Thence, the tensor products of the color octet with isospin states visualized in Fig.~\ref{fig2} are consolidated into two sixteen dimensional chiral irreps $(\mathbf{2}\otimes\mathbf{1} \otimes\mathbf{8})$ and $(\mathbf{1}\otimes\mathbf{2}\otimes\mathbf{8})$. There are now six new intermediary gauge bosons (the black roots) connecting fermion pairs of different helicity along the edges, explaining the curious feature of the cubic grouping described above. Collectively labeled $X$, three of the $X$ bosons are seen to carry color charge $r, g$, or $b$, electric charge $-\frac{1}{3}$, and weak hypercharge $-\frac{2}{3}$, whereas the other three have opposite charges (all are thus confined too). We denote the first three as $X^-_{r}, X^-_{g}, X^-_{b}$ and the second three as  $X^+_{\overline{r}}, X^+_{\overline{g}}, X^+_{\overline{b}}$. Like the $Y$ bosons, $X$ bosons carry charges that are integer multiples of $\frac{1}{3}$. 

By enlarging the symmetry of the color octet, all the color states have become interconnected along the cube's face diagonals and edges so that the fermions occupying them are consistently interconvertible within the cubic grouping.  The thirty-two fermions of one generation are, thus, described by the direct sum rep $(\mathbf{2}\otimes\mathbf{1} \oplus\mathbf{1}\otimes\mathbf{2})\otimes\mathbf{8}$, which is thirty-two dimensional.

\section{The discrete flavor symmetry group}
The cube embodying irrep $\mathbf{8}$, visualized in Fig.~\ref{fig3}, has the discrete symmetry group $S_4 \times \mathbb{Z}_2$, which is of order $24\times 2=48$. The nonabelian Symmetric group $S_4$ corresponds to the rotational symmetry or rigid motions of the cube, while $\mathbb{Z}_2$ is its reflection or parity symmetry. $\mathbb{Z}_2$ interchanges the diametrically opposite colors of the octet and, hence, is a symmetry only if accompanied by charge conjugation. Therefore, the discrete symmetry should be restricted to the Symmetric subgroup $S_4$. Thus, the gauge symmetry $so(7)$ entails Symmetric group $S_4$ for the color octet and, hence, for the fermion flavors shown below. 

$S_4$ has five irreps, namely, the trivial $\underline{\mathbf{1}}$, the alternating trivial $\underline{\mathbf{1}}^\prime$, the standard $\underline{\mathbf{3}}$, the alternating standard $\underline{\mathbf{3}}^\prime$, and the irrep $\underline{\mathbf{2}}$ which is the standard rep of $S_3$ induced on $S_4$ \cite{Fulton} (underlines distinguish the reps from those of the gauge group). The action of $S_4$ on the cube can be realized by the permutation of its four long diagonals. Thus, its action on the  vertices (color states) yields its eight-dimensional rep, which decomposes as
$\underline{\mathbf{8}}=\underline{\mathbf{1}}\oplus\underline{\mathbf{1}}^\prime\oplus\underline{\mathbf{3}}\oplus\underline{\mathbf{3}}^\prime$. In other words, concomitantly to the irrep $\mathbf{8}$ of the gauge algebra $so(7)$, the color octet is described by the reducible rep $\underline{\mathbf{8}}$ of the group $S_4$. The latter decomposes the colors in the octet by grouping them into four different flavor classes, such that $\underline{\mathbf{1}}\equiv w,\ \underline{\mathbf{1}}^\prime\equiv \overline{w}, \ \underline{\mathbf{3}}\equiv (r,g,b),\ \underline{\mathbf{3}}^\prime \equiv (\overline{r},\overline{g},\overline{b})$.  This is the flavor grouping already established by the cubic structure seen in Fig.~\ref{fig2}, which is reconciling since $S_4$ is implied by $so(7)$. Thus, the flavour symmetry group is not independent of the gauge group. 

In the literature, the Symmetric group $S_4$ has been conjectured as the flavor group independently of the gauge group  \cite{Ma2004,Hagedorn2006,SO10S4,Ishimori,MA2006,flavor}, leading to the quark-lepton complementary relation as well as realistic mixing matrices and mass hierarchies \cite{MA2006,flavor,mixing,mixing2}.

\section{The relationship with eleven dimensional spacetime}
Because $SU(2)_L\times SU(2)_R\cong Spin(4)$, the rep $(\mathbf{2}\otimes \mathbf{1}\oplus \mathbf{1}\otimes \mathbf{2})$ of $SU(2)_L\times SU(2)_R$ is the four dimensional spinor rep ${\mathfrak S}_4={\mathfrak S}_2^-\oplus {\mathfrak S}_2^+$ of $Spin(4)$, where ${\mathfrak S}_2^-\equiv \mathbf{2}\otimes \mathbf{1}, {\mathfrak S}_2^+\equiv \mathbf{1}\otimes \mathbf{2}$ are its half-spinor reps. Hence, the left/right isospin states are represented by  ${\mathfrak S}_2^\mp$ and the direct sum rep of the thirty two fermions can be expressed as $\mathfrak{S}_4 \otimes \mathbf{8}$. Now,  the noncompact (double cover) Lorentz group, $Spin(3,1)\cong SL(2,\mathbb{C})$, has the same  algebra as the compact $Spin(4)$, namely $so(4)$. This isomorphism in algebra defines a  correspondence between  the LC/RC half-spinor  (Weyl) reps of $Spin(3,1)$ and the ${\mathfrak S}_2^\mp$ of $Spin(4)$, which gives the isospin states their chirality. 

As a corollary, the algebra of the group $SO^+(3,1)\times Spin(7)\subset Spin(10,1)$ is isomorphic to the proposed GUT symmetry algebra. This group represents the symmetry of an eleven-dimensional spacetime with a decomposable structure due to the product form of the group. The decomposition is into two separate parts, namely the usual Riemannian spacetime $M^4$ with local Lorentz symmetry  $SO^+(3,1)$, and a seven sphere $S^7$ which compactifies the remaining seven spatial dimensions. $S^7$ is the manifold of unit octonions whose tangent space at each point, spanned by imaginary octonions,  has the rotation symmetry group $Spin (7)$. Being associated with octonions, $S^7$ is not a group manifold but is nevertheless parallelizable. It is the unique compact, simply connected non-group manifold that is parallelizable. The subgroup of $Spin(7)$ which fixes a point on $S^7$ (the stabilizer of any one point) is the exceptional group $G_2$, the automorphism group of octonions. $G_2$ has been considered in relation to compactification in M-theory \cite{G2}, while both groups are of interest in the context of special geometric structures \cite{old}.

\begin{figure}
\centering
{\includegraphics[width=6cm]{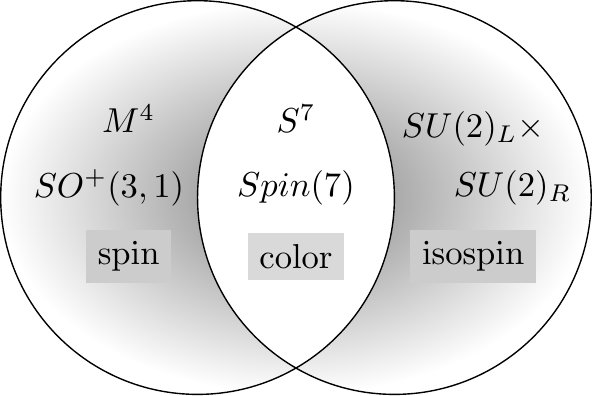}}
\caption{Spacetime and gauge symmetries. The left circle represents the symmetry group of the eleven-dimensional spacetime $M^4\times S^7$, while the right circle is the gauge group of the elementary particles.  $Spin(7)$, which belongs to both symmetry groups, lies in the intersection of the two circles. Spin states are associated with $M^4$ and its Lorentz symmetry, color states with $S^7$ and its $Spin(7)$ symmetry, and isospin states with the  $SU(2)_L\times SU(2)_R$ gauge symmetry.} \label{fig4}
\end{figure}

The correspondence between the GUT gauge group and the symmetry group of the spacetime $M^4\times S^7$ suggests that the latter is apt to describe elementary particles and interactions. Thence, the gauge group $Spin(7)$ is also a spacetime symmetry group, and its spinor rep $\mathbf{8}$ which describes the color states, is a $S^7$ spinor.  On the other hand, the Weyl irreps of the Lorentz group, which represent spin states, are $M^4$ spinors, while isospin states are associated with the gauge group $SU(2)_L\times SU(2)_R$. The situation is picturized in Fig.~\ref{fig4}. 

\section{Spontaneous symmetry breaking of the model}
The compact submanifold, $S^7$, is not realizable at low energies, which implies that the $Spin(7)$ symmetry must break. This occurs in the SSB 
into the (unbroken) SM, which also breaks the parity or LR symmetry, as
 $SU(2)_R\times Spin(7)\rightarrow  SU(3)_C \times U(1)_Y.$
(This is akin to the PS model where $SU(4)$ replaces $Spin(7)$. One may also envisage an intermediate LR symmetric step involving the $B-L$ hypercharge, namely $ SU(2)_R\times SU(3)_C \times U(1)_{B-L}$, like in the PS model.) The EW symmetry breaks at a much lower energy scale ($\sim 10^2$ GeV). As in LR symmetric models \cite{LS} in general, there can be several ways to implement these SSBs, differing in the choice of the scalar (Higgs) sector. However, as a common feature, the scalar sector must always contain a bidoublet Higgs field for Dirac mass to exist in the Lagrangian, thus yielding tree-level mass. A real bidoublet forces mass equality between isospin partners at tree level, an issue which is usually avoided by choosing a complex bidoublet, akin to Two Higgs Doublet Models \cite{Volkas}. A simple choice of the scalar sector that works is then to add a right Higgs multiplet in the standard rep {\bf 7} of Spin(7). The superheavy multiplet is responsible for SSB to the SM gauge group, whereas the light bidoublet Higgs breaks the EW symmetry and gives tree-level mass to Dirac fermions.  

In this model of the scalar sector, the superheavy multiplet Higgs field $\phi$ is represented by $(\mathbf{1}\otimes \mathbf{2}\otimes \mathbf{7})$ and the light  complex bidoublet by $(\mathbf{2}\otimes \mathbf{2}\otimes \mathbf{1})$. We, therefore, expect twenty-eight superheavy and eight light Higgs particles. The SSB into SM yields $24-9=15$ broken generators, resulting in fifteen superheavy gauge bosons, which are the $X$ and $Y$ particles and the $Z^0_R,\ W^\pm_R$. Their mass comes from ``eating'' fifteen of the twenty-eight Higgs particles, leaving $28-15=13$ superheavy Higgs bosons. The remaining $27-15=12$ massless gauge bosons are precisely those of the unbroken SM gauge group. When this group finally breaks, it yields $12-9=3$ broken generators and, hence, the light gauge bosons  $Z^0,\ W^\pm$. Their mass comes from eating three of the eight Higgs particles leaving $8-3=5$ light Higgs bosons.

Let us explicitly demonstrate the above particle spectrum for the first SSB as an example, the second (EW SSB of SM) being quite standard \cite{Volkas, Barbensi}. $\phi$ and $\chi$ are $(2\times 7)$ and $ (2\times 2)$ matrices that transform under the gauge group as $\phi\rightarrow U_R\phi U_7^T$ and $\chi \rightarrow U_L \chi U_R^\dagger$, where $U_7,U_{R,L}$ belong to $SO(7), SU(2)_{R,L}$, respectively. The most general scalar potential allowed by the gauge symmetry and renormalizability constraint is
$$
V(\phi, \chi)=V_1(\phi)+V_2(\chi)+V_3(\phi,\chi)
$$
where
\begin{eqnarray}
 V_1(\phi)=&-&\mu_1^2 Tr(\phi \phi^\dagger)+\beta_1 [Tr(\phi\phi^\dagger)]^2+\beta_2 Tr[(\phi\phi^\dagger )^2]\nonumber \\
 V_2(\chi)=&-&\mu_2^2 Tr(\chi^\dagger \chi)+\eta_1 [Tr(\chi^\dagger \chi)]^2+\eta_2 Tr[(\chi^\dagger \chi)^2]\nonumber\\
 &-&\mu_3^2 \left( Tr(\chi^\dagger \tilde\chi)+h.c\right) 
+ \eta_3\left( [Tr(\chi^\dagger \tilde\chi)]^2+h.c\right)+\eta_4 \left(Tr[(\chi^\dagger \tilde\chi)^2]+h.c\right)\nonumber \\
V_3(\phi, \chi)=&\ &\xi_1Tr(\phi \phi^\dagger)Tr(\chi^\dagger \chi)+\xi_2Tr(\phi \phi^\dagger\chi^\dagger \chi)+\xi_3 Tr(\phi \phi^\dagger\tilde\chi^\dagger \tilde\chi)\nonumber\\
&+&\xi_4Tr(\phi \phi^\dagger)\left( Tr(\chi^\dagger \tilde\chi)+h.c\right) 
+\xi_5 \left(Tr(\phi \phi^\dagger\chi^\dagger \tilde\chi)+h.c \right)\nonumber
\end{eqnarray}
with $\tilde\chi=\tau_{2L} \chi^* \tau_{2R}$, $\tau_2$ being the Pauli matrix. In above, $\mu_1, \mu_2,\mu_3$ have mass dimension one and all other coefficients are dimensionless. The first SSB involves $\phi$, which initially has twenty-eight (massive) degrees of freedom as seen in the mass term of $V_1(\phi)$. The SSB to SM occurs when $\phi$ takes the following vacuum expectation value 
$$
\phi_0=
\begin{pmatrix}
0&0&0&0&0&0&\frac{v}{\surd 2}
\\
0&0&0&0&0&0&0
\end{pmatrix}
$$
to minimize the potential, where $v=\mu_1/\sqrt{\beta_1+\beta_2}$, giving $V_1(\phi_0)=-\mu_1^2v^2/4$. Expanding the potential in the small neighborhood of $\phi_0$ yields
\begin{eqnarray}
V_1(\phi)&=&V_1(\varphi+\phi_0)\nonumber\\
&=&V_1(\phi_0)-\beta_2v^2\left(|\varphi_{21}|^2+|\varphi_{22}|^2+|\varphi_{23}|^2+|\varphi_{24}|^2+|\varphi_{25}|^2+|\varphi_{26}|^2+2[Re(\varphi_{17})]^2\right)\nonumber
\end{eqnarray}
which corroborates that only thirteen (massive) degrees of freedom remain after SSB, as anticipated (note the conditions $\beta_2, \beta_1+\beta_2>0$). Those disappearing, give their masses to the fifteen goldstone bosons (broken generators) to create the superheavy gauge particles.

 \section{Physical consequences}
 In contrast to the PS model, the proposed extension predicts $B-L$ conserving nucleon decay $p\rightarrow \pi^0 e^+,\ n\rightarrow \pi^- e^+$, via $X$-boson exchange as in other GUTs. For example, a red $u$ quark in the proton $uud$ can become a positron (of opposite helicity) by emitting a $X^-_r$ boson. This boson could be absorbed by a blue $d$ quark, converting it into a green $\overline{u}$ quark (of opposite helicity), which would then combine with the remaining green $u$ quark of the proton to form a $\pi^0$ meson. In the case of neutron $udd$, the only difference is the remaining quark, which would be a green $d$ quark, thus yielding a $\pi^-$ meson instead. Such decays may be suppressed in theories with extra dimensions ($S^7$ in our case) to comply with the experimental lower bound on proton's half-life ($\sim 10^{-34}$ yr) \cite{Proton}. Thus, the nucleon decay prediction and control have furnished a good testing ground for GUTs and it is to be determined whether the proposed model passes this test.
 
Another consequence concerns lepto-quark mixing. In the PS model, the lepto-quark mixing mediated by the $Y$ bosons can break the lepton flavor universality \cite{LFU} at tree level, which is an accidental symmetry of the SM.  This would lead to new sources of CP violation through additional phases in the Yukawa couplings, deemed necessary for explaining matter-antimatter asymmetry in the universe. The proposed extension enhances this mechanism by the extra lepto-quark mixing via the $X$ bosons.

\end{document}